\PassOptionsToPackage{table,xcdraw}{xcolor}
\documentclass[conference,a4paper]{APSIPA2025}
\usepackage{amsmath,amssymb}
\usepackage{graphicx}
\usepackage{multirow}
\usepackage{threeparttable}

\usepackage[style=ieee,backend=biber]{biblatex}
\usepackage{balance}

\usepackage{booktabs}
\usepackage{soul}

\usepackage{geometry}
\usepackage{fancyhdr}

\fancypagestyle{firststyle}{
  \fancyhf{}
  \fancyhead[C]{2025 Asia Pacific Signal and Information Processing Association Annual Summit and Conference (APSIPA ASC)}
}

\begin{document}

\title{RawTFNet: A Lightweight CNN Architecture for Speech Anti-spoofing}

\author{
\authorblockN{
Yang Xiao\authorrefmark{1},
Ting Dang\authorrefmark{1},
Rohan Kumar Das\authorrefmark{2}
}

\authorblockA{
\authorrefmark{1}The University of Melbourne, Australia}

\authorblockA{
\authorrefmark{2}Fortemedia Singapore, Singapore}
}

\maketitle
\thispagestyle{firststyle}
\pagestyle{fancy}


\begin{abstract}
Automatic speaker verification (ASV) systems are often affected by spoofing attacks. Recent transformer-based models have improved anti-spoofing performance by learning strong feature representations. However, these models usually need high computing power. To address this, we introduce RawTFNet, a lightweight CNN model designed for audio signals. The RawTFNet separates feature processing along time and frequency dimensions, which helps to capture the fine-grained details of synthetic speech. We tested RawTFNet on the ASVspoof 2021 LA and DF evaluation datasets. The results show that RawTFNet reaches comparable performance to that of the state-of-the-art models, while also using fewer computing resources. The code and models will be made publicly available.
\end{abstract}

\section{Introduction}

In recent years, deep learning technologies have greatly improved speech generation methods such as text-to-speech (TTS)~\cite{blizzard2023} and voice conversion (VC)~\cite{VCC2020}. These technologies bring many advantages, especially for speech applications. However, they also create risks when used to generate fake speech that imitates a target speaker. Modern TTS and VC systems can now produce speech that sounds highly natural, which makes it easier to attack voice-based systems. As a result, automatic speaker verification (ASV) systems face growing threats from such synthetic speech~\cite{survey1,asv,survey2}. To reduce these risks, researchers have focused on anti-spoofing methods that can detect fake voices. The ASVspoof challenge~\cite{ASVspoof_journal} has become an important platform in this field. It provides standard datasets, rules, and evaluation methods to help the research community develop and compare anti-spoofing systems.

Deep learning-based speech anti-spoofing methods can be divided into two main types: feature-based and end-to-end approaches. Feature-based methods use pre-designed representations such as linear-frequency cepstral coefficients (LFCC)~\cite{comparison}, cochlear filter cepstral coefficient and instantaneous frequency~\cite{PatelINTERSPEECH2015} and constant-Q transform (CQT) based representations~\cite{constant,rkd_is2019}. These features help to capture the unique patterns of fake speech for detection. 

On the other hand, end-to-end approaches work directly with raw audio signals~\cite{rawbmamba,rawboost,xlsrmamba}. They do not depend on handcrafted features and have shown strong generalization and performance. As a result, many recent studies focus on end-to-end models. For example, RawNet2~\cite{rawnet2} uses time-domain convolution to learn features from raw waveforms. Attention-based models, like AASIST~\cite{aasist}, apply graph attention networks to better capture both local and global speech features across time and frequency. Similarly, Rawformer~\cite{rawformer} combines convolutional layers with transformers, using self-attention to detect important clues for fake speech. Although these models perform well, they often need large computing resources. This makes them hard to use in real-world systems, especially where low power and fast processing are required. Therefore, it is important to find a balance between accuracy and efficiency for practical deployment.

To improve efficiency in speech processing tasks, researchers have proposed several lightweight convolutional neural network (CNN) architectures. For example, BC-ResNet~\cite{broadcasted} applies depthwise separable convolutions to reduce both model size and computational cost, while maintaining strong performance. Similarly, the TDNN~\cite{tdnn} captures temporal structures in speech by applying convolution over time steps. These models are widely used in tasks such as ASV due to their good balance between speed and accuracy. However, although these architectures are efficient, they are not specifically tailored for speech anti-spoofing. Most of them concentrate on temporal features or use global representations, without focusing on the unique characteristics of synthetic speech. In practice, fake speech signals often contain subtle but important artifacts that are distributed across both time and frequency dimensions~\cite{tf}. Standard CNNs, which process the input uniformly or emphasize only one dimension, may overlook these subtle patterns.

To handle the challenges of speech anti-spoofing, a task-specific model that can separate time and frequency information is needed. Inspired by time-frequency attention~\cite{cai2024tf}, we propose Time-Frequency Convolutions (TF-Convs), a 1D convolution-based module. It processes feature maps in two branches: one for time and another for frequency. This allows the model to capture detailed patterns in both dimensions while keeping computation low. We further propose a new anti-spoofing architecture (RawTFNet) that combines both efficiency and task-specific design. First, we use a Sinusoidal Convolution~\cite{sinc} layer and some SE-Res2Net blocks as the frontend, which effectively captures local patterns and enhances feature diversity. Then, the resulting feature map is processed by the TF-Convs module. Finally, the combined features are passed into the classifier. This architecture is designed to detect the fine-grained artifacts while keeping the model efficient. We tested RawTFNet on the  ASVspoof 2021 LA and DF datasets. The results show that RawTFNet works directly on raw features and reaches comparable accuracy to state-of-the-art (SOTA) models, while using only 70K parameters.

\begin{figure*}[t!]
    \centering
    \includegraphics[width=0.9\linewidth]{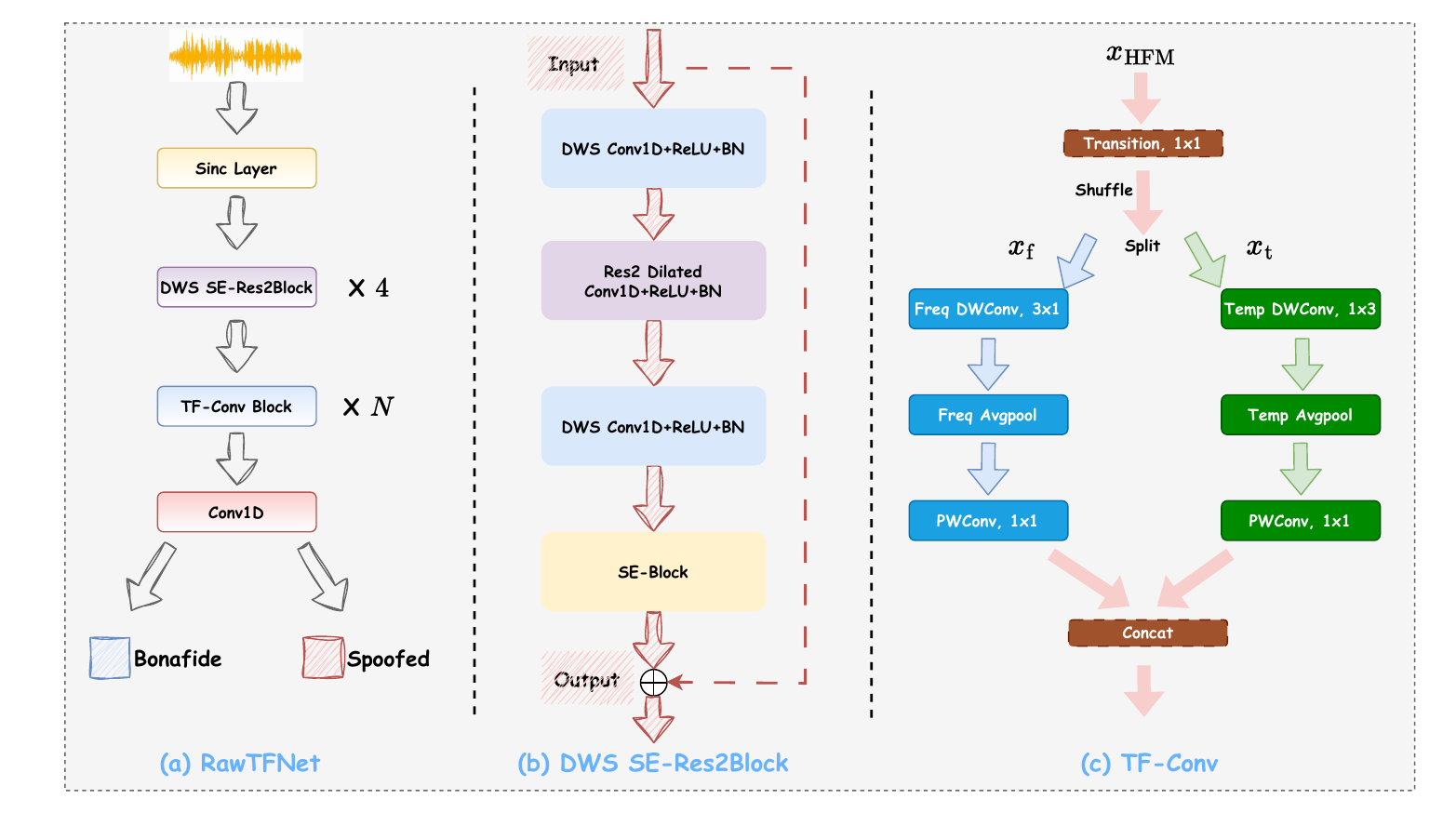}
    \vspace{-5mm}
    \caption{ Overview of the proposed RawTFNet architecture for speech anti-spoofing. (a) The model begins with a Sinc layer followed by four DWS SE-Res2Blocks to extract low-level features. These are then processed by stacked TF-Conv blocks and a final Conv1D layer for classification.  (b) The DWS SE-Res2Block integrates depthwise separable convolutions, a Res2Net module with dilation, and a squeeze-and-excitation block to enhance channel-wise attention and efficiency. (c) The TF-Conv module separates feature maps along time and frequency dimensions using 1D depthwise convolutions, followed by pooling and pointwise convolution before merging. We fix the TF-Conv layer number `N' to 9 in this work.}
    \label{fig:enter-label}
\end{figure*}

\section{Proposed RawTFNet}

To address the limitations of existing lightweight speech anti-spoofing models that do not explicitly separate time and frequency information, we propose a task-specific architecture for anti-spoofing called RawTFNet. Inspired by the time-frequency nature of speech signals, RawTFNet is designed to better capture fine-grained artifacts using dimension-aware processing. Fig.~\ref{fig:enter-label} (a) illustrates the overall structure of the proposed RawTFNet. In particular, we introduce a novel TF-Convs module that separates the analysis of time and frequency using 1D depthwise convolutions. Additionally, to improve the representation of local and contextual information, we integrate DWS SE-Res2Blocks that enhance feature diversity and channel attention with low computation.

\subsection{Frontend}
The frontend in RawTFNet is designed to extract low-level spectro-temporal features directly from raw audio. We adopt a modified version of RawNet2~\cite{rawnet2} to perform this task. First, the raw waveform is passed through a set of trainable Sinc filters~\cite{sinc} (with 70 filters in total), which act as learnable band-pass filters. This converts the signal into a two-dimensional feature map of frequency and time bins.

Next, the extracted features are refined through a series of convolutional layers. The resultant output passes through a ResNet block with 32 filters to capture early patterns. Then, it is processed by three SE-Res2Net blocks with 64 filters each, which enhance feature diversity using multiple receptive fields. To further improve efficiency, we replace standard 2D convolutions with DWS convolutions. This leads to the use of DWS SE-Res2Net blocks, as shown in Fig.~\ref{fig:enter-label} (b), which reduce computation while preserving performance. Each block also includes a squeeze-and-excitation (SE) module~\cite{hu2018squeeze} to strengthen important channel-wise information. As a result, the encoder generates high-level feature maps \(x_{\text{HFM}} \in \mathbb{R} ^{C\times F\times T}\), where \(C\), \(F\), and \(T\) represent the number of channels, frequency bins, and time steps, respectively.

\subsection{Time-Frequency Convolutions}
To better detect the subtle patterns of synthetic speech from the high-level feature maps, we introduce Time-Frequency Convolutions (TF-Convs), a novel module designed specifically for anti-spoofing tasks. Unlike conventional CNN layers that treat all dimensions equally, TF-Convs separates the processing of time and frequency to capture finer details that are often overlooked. This design enables the model to focus on spoofing-related cues that may appear in different forms along the two axes. 

As shown in Fig.~\ref{fig:enter-label}, TF-Convs start with a \(1\times1\) transition convolution, which adjusts the number of channels from \(C\) to \(C'\), transforming the input \(x_{\text{HFM}} \in \mathbb{R} ^{C\times F\times T}\) into \(\mathbb{R} ^{C'\times F\times T}\). Then, a channel shuffle operation is applied to enhance feature mixing. The resulting tensor is evenly split along the channel dimension into two parts: one for the frequency path \(x_f\) and one for the time path \(x_t\), each of shape \(\mathbb{R} ^{C'/2 \times F\times T}\). Each branch is processed separately. The frequency path uses a \(3\times1\) depthwise convolution, followed by frequency-wise average pooling and a \(1\times1\) pointwise convolution. Similarly, the time path applies a \(1\times3\) depthwise convolution, followed by time-wise pooling and a pointwise convolution. These operations produce compact 1D features \(v_f\) and \(v_t\). 

Next comes the broadcasting operation~\cite{broadcasted} that copies a 1D feature (such as \(v_f\)) across one dimension to match the size of the original 2D feature map (e.g., \(F\times T\)). This operation allows the model to add the compact 1D summary back to the full feature map in a way that maintains structure but reduces computation. The updated maps \(\hat{x}_f\) and \(\hat{x}_t\) are then created using element-wise addition between the original split maps and their corresponding broadcasted vectors, as shown in equations~\eqref{eq:xhat_f} and~\eqref{eq:xhat_t}.

\begin{equation}\label{eq:xhat_f}
\hat{x}_{f}
\;=\;
\sum_{j=1}^{T}\bigl(x_{ij,f} + v_{j,f}\bigr)
\end{equation}

\begin{equation}\label{eq:xhat_t}
\hat{x}_{t}
\;=\;
\sum_{i=1}^{F}\bigl(x_{ij,t} + v_{i,t}\bigr)
\end{equation}

Finally, the two enhanced outputs are concatenated along the channel axis to form the final output \(y \in C'\times F\times T\). All layers include batch normalization and ReLU activation to improve training stability. By separating and refining the time and frequency information, TF-Convs help the model learn more discriminative features for detecting fake speech.
\subsection{RawTFNet Architecture}

In summary, the architecture of RawTFNet is designed to achieve strong spoofing detection accuracy with low computational cost. It is composed of three main components: a SinConv and DWS SE-Res2Net frontend, a stack of TF-Convs modules, and a Conv1D classifier. The frontend begins by extracting low-level spectro-temporal features from raw audio using a learnable Sinc filter, followed by DWS SE-Res2Blocks that enhance both local patterns and channel-wise information. These early layers allow the model to capture fine-grained acoustic cues that are critical for detecting synthetic speech.

After encoding, the features are refined through nine TF-Convs modules, each of which processes time and frequency dimensions separately. This separation allows the model to detect subtle artifacts in both directions. To improve abstraction, two \(2\times2\) max-pooling layers with a stride of 2 are inserted for intermediate downsampling. The final output is fed into a lightweight classifier composed of global average pooling, a 1D convolution layer, which outputs a binary score for bonafide or spoofed speech. To support different computing environments, a channel width parameter \(\tau\) is used to control the model’s complexity. This makes RawTFNet-\(\tau\) flexible and suitable for both edge devices and high-performance systems.

\section{Experimental Setup}

\subsection{Datasets}
To evaluate the effectiveness and generalization ability of the proposed RawTFNet, we conducted experiments on three benchmark datasets: ASVspoof 2021 LA, ASVspoof 2021 DF, and In-the-Wild. All models were trained using the ASVspoof 2019 LA training set~\cite{asvspoof2019}, which provides labeled bonafide and spoofed speech samples for supervised learning.

The ASVspoof 2021 LA evaluation set contains about 180,000 utterances, including both real and spoofed speech generated through advanced TTS and VC systems~\cite{asvspoof2021}. The ASVspoof 2021 DF set consists of approximately 600,000 utterances, where audio samples are compressed using lossless codecs, reflecting real-world media storage conditions. 

\subsection{Metrics}
We used two standard metrics to measure anti-spoofing performance: the Equal Error Rate (EER) and the minimum tandem detection cost function (min t-DCF)~\cite{tDCF_Tomi_IEEE}, both widely adopted in the ASVspoof Challenges. To assess model efficiency, we also report two system-level metrics: Parameters (PARAMS), which denotes the total number of model parameters, and MACS, referring to the number of million multiply-accumulate operations to measure computational cost.

\subsection{Implementation Details}

For training, we follow the baseline setup introduced in~\cite{rawnet2}, using the ASVspoof 2019 LA dataset~\cite{asvspoof2019}. Each audio sample is either cropped or concatenated to create a fixed-length input segment of approximately 4 seconds (64,000 samples). We adopt the Adam optimizer, with a weight decay of \(1\times10^{-4}\) to prevent overfitting. The model is trained using a batch size of 32 and a learning rate of \(1\times10^{-4}\).

As the loss function, we use weighted cross-entropy, which accounts for class imbalance between bonafide and spoofed samples. Following~\cite{rawboost}, we apply RawBoost augmentation, specifically the `series: (1+2+3)' combination (Algo4), to improve model robustness. All models are trained for 100 epochs, and for evaluation, we average the top five checkpoints based on validation performance to reduce variability and improve stability during testing.

\begin{table}[t]
\centering
\caption{Comparison of SOTA single anti-spoofing systems on ASVspoof 2021 LA and DF evaluation sets. }
\vspace{-2mm}
\label{tab:sota}
\resizebox{\columnwidth}{!}{%
\begin{tabular}{lccccc}
\toprule
                          \multirow{2}{*}{\textbf{Systems}} &   \multirow{2}{*}{\textbf{PARAMS}} &   \multirow{2}{*}{\textbf{MACS}}   & \multicolumn{2}{c}{\textbf{2021LA}} & \textbf{2021DF} \\ \cmidrule{4-6} 
  & & & EER\%               & min t-DCF               & EER\%                 \\ \midrule

 RawGT-ST~\cite{rawgt}               & 0.44M    & 17.4G           & 10.25                & -                   & 23.26                    \\
 RawNet2~\cite{rawnet2}               & 17.60M    & \textbf{0.8G}           & 9.50                & -                   & 22.38                    \\
  AASIST~\cite{aasist}               & 0.30M    & 8.9G           & 10.51                 & 0.488                   & 21.07                    \\
 SE-Rawformer~\cite{rawformer}         & 0.37M  & 6.1G             & 4.98                & 0.318                   & 20.26                         \\
 Rawformer-L~\cite{rawformer}           & 0.29M & 9.2G                & 6.83                & 0.361                   & -                    \\ \midrule
 \cellcolor[HTML]{C4D5EB}RawTFNet-16 (Proposed)    & \cellcolor[HTML]{C4D5EB}\textbf{0.07M}  & \cellcolor[HTML]{C4D5EB}2.9G         & \cellcolor[HTML]{C4D5EB}\textbf{4.50}                & \cellcolor[HTML]{C4D5EB}\textbf{0.295}                   & \cellcolor[HTML]{C4D5EB}18.53 \\ 
\cellcolor[HTML]{C4D5EB}RawTFNet-32 (Proposed)    & \cellcolor[HTML]{C4D5EB}0.17M  & \cellcolor[HTML]{C4D5EB}5.4G         & \cellcolor[HTML]{C4D5EB}5.05                & \cellcolor[HTML]{C4D5EB}0.321                   & \cellcolor[HTML]{C4D5EB}\textbf{16.82}                      \\ \bottomrule
\end{tabular}%
}
\vspace{-4mm}
\end{table}

\section{Results and Discussions}

\subsection{Comparison Study }
Table I presents a comparison between the proposed RawTFNet-16 and RawTFNet-32 models and several SOTA anti-spoofing systems on the ASVspoof 2021 LA and DF evaluation sets. Both variants of RawTFNet demonstrate a strong balance between detection performance and model efficiency. In terms of model size, RawTFNet-16 has only 0.07M parameters and 2.9G MACs, making it the most lightweight model in the comparison. RawTFNet-32 is slightly larger at 0.17M parameters and 5.4G MACs, but still significantly smaller and more efficient than models like RawNet2 (17.60M parameters with fewer MACs due to its heavy FC layers), AASIST (0.30M, 8.9G), and Rawformer-L (0.29M, 9.2G). These results indicate that both RawTFNet variants are highly suitable for resource-constrained and real-time environments, with RawTFNet-16 offering an extremely compact footprint.

In terms of detection performance, RawTFNet-32 achieves an EER of 5.05\% and min t-DCF of 0.321 on the ASVspoof 2021 LA set. This is on par with SE-Rawformer (4.98\%) but with much lower complexity. Meanwhile, RawTFNet-16 achieves an even lower EER of 4.50\% and min t-DCF of 0.295, suggesting that smaller models can still maintain or even improve accuracy if designed with effective feature processing mechanisms. This performance surpasses larger models such as AASIST (EER 10.51\%, min t-DCF 0.488), demonstrating the benefit of RawTFNet’s task-specific architecture. On the more challenging ASVspoof 2021 DF dataset, both RawTFNet variants again show strong generalization. RawTFNet-32 achieves the lowest EER of 16.82\% among all systems, outperforming RawNet2 (22.38\%), AASIST (21.07\%), and SE-Rawformer (20.26\%). RawTFNet-16 also performs well with an EER of 18.53\%, showing that even the smallest version of our model generalizes effectively to unseen, codec-distorted speech.

Overall, these results confirm the strength of RawTFNet’s time-frequency separation strategy, depthwise convolutional design, and flexible scaling. Both RawTFNet-16 and RawTFNet-32 offer a compelling trade-off between performance and efficiency, making them suitable for diverse real-world deployments.

\begin{figure}
    \centering
    \includegraphics[width=\linewidth]{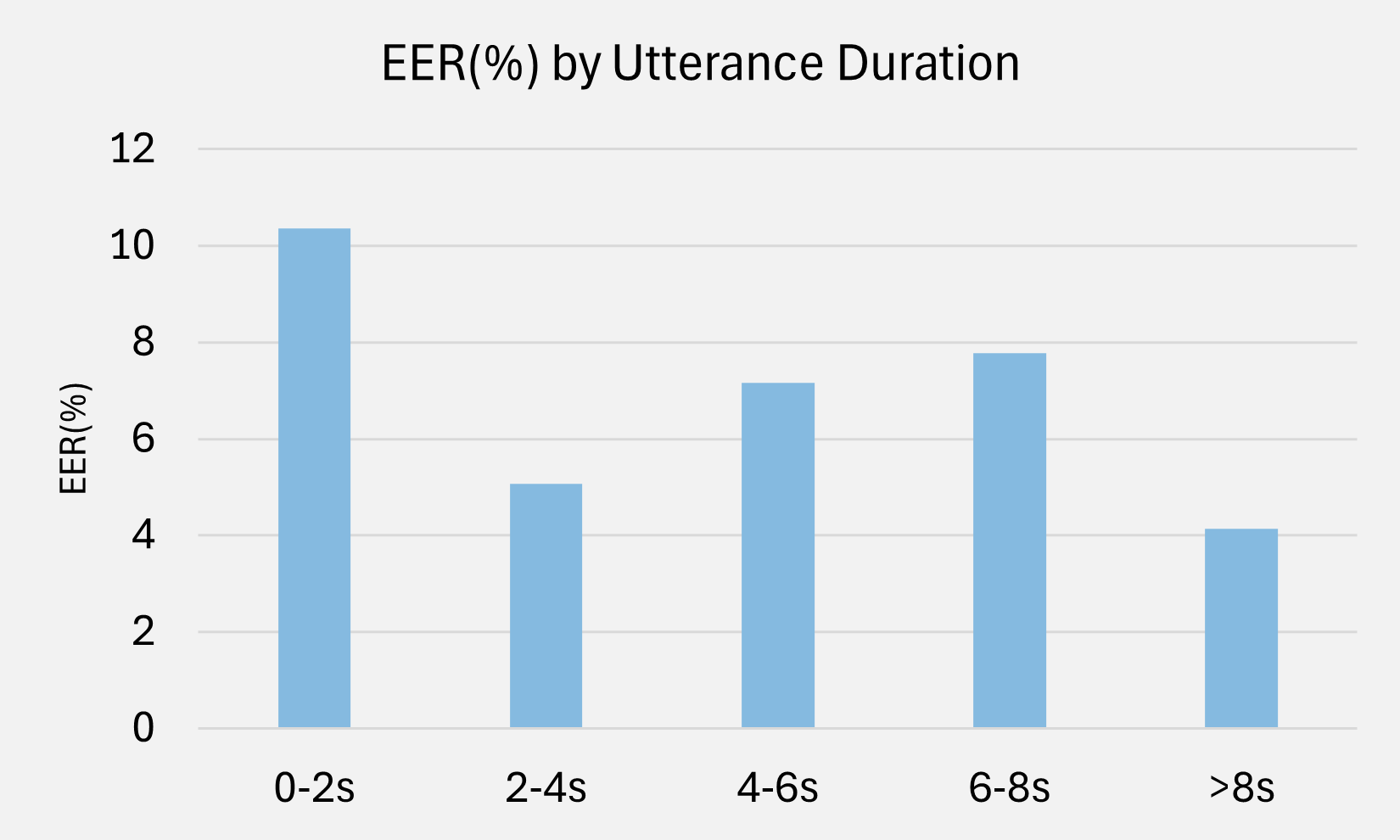}
    \caption{EER (\%) across different utterance duration ranges on the ASVspoof evaluation set. Each bar represents a specific duration group, and `$n$’ indicates the number of utterances in that group.}
    \label{fig:2}
\end{figure}

\subsection{Impact of Utterance Duration on Detection Performance}
Fig.~\ref{fig:2} shows how the EER of RawTFNet-32 varies across different utterance durations. The model performs worst on very short clips (0–2s), with EER above 10\%, likely due to insufficient acoustic context. Performance improves significantly for 2–4s inputs, reaching the lowest EER ($\sim$5\%), matching the model’s training segment length. However, EER slightly increases for 4–6s and 6–8s ranges, possibly due to added variability or training bias. Interestingly, EER drops again for utterances longer than 8s, though this is based on fewer samples. These results suggest that utterance length influences spoof detection accuracy and highlight the potential of duration-aware training strategies.

\begin{table}[t]
  \centering
  \begin{threeparttable}
    \caption{Ablation Study on RawTFNet-32. `w/o’ indicates removal of the specified component from the full model. EER (\%) is reported on the ASVspoof 2021 LA eval set. PARAMS (K) and MACs (M) represent the number of parameters and Multiply–Accumulate Operations, respectively.}
    \label{tab:ablation_rawtfn32}
    \begin{tabular}{lccc}
      \toprule
      Model Variant        & EER (\%) & MACs (M) & Param (K) \\ 
      \midrule
      RawTFNet-32 (Full)   & \textbf{5.05}     & 5400     & 170       \\
      w/o freq branch      & 6.91     & 4550     & 150       \\
      w/o temp branch      & 7.08     & 4480     & 150       \\
      w/o shuffle          & 5.74     & 5400     & 170       \\
      \bottomrule
    \end{tabular}
  \end{threeparttable}
\end{table}

\subsection{Ablation Study}

To analyze the impact of each component in RawTFNet-32, we perform an ablation study on the ASVspoof 2021 LA set, as shown in Table II. Removing the frequency or temporal branch significantly increases the EER to 6.91\% and 7.08\%, respectively, confirming the importance of modeling both time and frequency information. Excluding the shuffle operation leads to a moderate drop in performance (EER: 5.74\%) without affecting model size or cost, showing its positive role in feature interaction. These results demonstrate that all components contribute to RawTFNet’s effectiveness and efficiency.

\section{Conclusions}
In this work, we introduced RawTFNet, a lightweight and effective model for speech anti-spoofing. By separating the processing of time and frequency features through the proposed TF-Convs, RawTFNet captures detailed cues in both dimensions while maintaining low computational cost. We further enhanced efficiency using DWS SE-Res2Blocks and a flexible channel width design. Experimental results on ASVspoof 2021 LA and DF evaluation datasets show that RawTFNet achieves competitive accuracy with fewer parameters and lower MACs compared to SOTA systems. Ablation studies confirm the importance of each component, especially the separate time-frequency paths. In summary, RawTFNet offers a robust and efficient framework, well-suited for deployment in real-world anti-spoofing scenarios.


\balance
\printbibliography

\end{document}